\documentclass[preprint,onecolumn,aps,amsmath,nofootinbib,superscriptaddress]{revtex4}

\usepackage{graphicx}
\usepackage{bm}
\usepackage{epsfig}
\usepackage{color}
\usepackage{verbatim} 

  \newcount\hour \newcount\minute
  \hour=\time \divide \hour by 60
  \minute=\time
  \newcommand{\mydate}{\ \today \ - \number\hour :\ifnum \minute<10 0\fi 
\number\minute}

\def\OMIT#1{}

\newcommand{\bea}{\begin{eqnarray}}
\newcommand{\eea}{\end{eqnarray}}

\newcommand{\beq}{\begin{equation}}
\newcommand{\eeq}{\end{equation}}

\begin{document}


\title{\boldmath
  Analytic Virtual Corrections for Higgs Transverse Momentum Spectrum  at $O(\alpha_s^2/m_t^3)$ via Unitarity Methods
}

\author{Duff Neill}
\affiliation{Department of Physics, Carnegie Mellon University,
    Pittsburgh, PA 15213  \vspace{0.3cm}}

\begin{abstract}
  
Utilizing the Higgs Effective Theory, including dimension eight operators, in conjuction with spinor-helicity and unitarity methods,  
I present analytic expressions for the virtual corrections for $gg\rightarrow Hg$ and $q\overline{q}\rightarrow Hg$
amplitudes at order $\alpha_s^2/m_t^3$. These contributions become enhanced as the invariant final state mass increases.
\end{abstract}

\maketitle

\section{Introduction}

The Higgs boson is the last remaining piece of the standard model to be observed, and its 
observation has been the impetus behind much theoretical and experimental work. The dominant
production mechanism at the LHC is through a gluon initiated top loop, and this mode of 
production has been well studied to high orders in both the top mass and strong coupling expansions. 
A different but related observable of interest is the Higgs produced with a large transverse momentum. 
This observable has received considerable attention \cite{Glosser:2002gm,Keung:2009bs,PhysRevD.67.073003},
both a probe of new physics \cite{ARZ} and as experimental avenue for Higgs discovery\cite{deFlorian:1999zd}. 
To produce a Higgs with a large transverse momentum, it is necessary for it to recoil off a hard 
jet. This makes it particularly difficult for study, as the gluon fusion process begins 
at one loop, and higher order QCD corrections to the Higgs $p_t$ are two-loop four-point 
functions. Practically, one can either tackle the higher order full theory contributions through 
asymptotic expansions \cite{Harlander:2009mq,Harlander:2009bw,Pak:2009bx}, 
or extend the effective theory, obtained from integrating out the top quark, to include more heavy mass 
effects \cite{Neill:2009tn}. Extending the effective theory is useful 
from several perspectives. Corrections in the effective theory are universal, in the sense that 
incorporating new heavy physics is a matter of adjusting matching coefficients. Further, 
by gauge invariance, once one has the matching coefficients, one can include higher order heavy 
mass and loop corrections to many other processes, including the phenomenologically important Higgs 
production in association with two jets, relevant for determination of Higgs coupling to
the weak vector bosons. All one needs is a one loop correction in the effective theory, 
where methods of loop calculations have become quite sophisticated. In this paper I present the 
one loop corrections to Higgs production with a hard jet in the extended Higgs Effective theory.
With these corrections one can reproduce the full theory amplitude to order $O(\alpha_s^2/m_t^3)$,
when combined with the matching coefficients for two-loop hard region.

\section{The Basis}
The operator basis used in calculating the matching coefficients in an earlier paper \cite{Neill:2009tn}, is

\begin{align*}
 O_1 = \frac{H F^{a}_{\mu\nu}F^{a \mu\nu}}{v} \; O_2 = \frac{H D_{\alpha}F^{a}_{ \mu\nu}D_{\alpha}F^{a \mu\nu}}{v M^2} \; O_3 = \frac{HF^{a\mu}_{\nu}F^{b\nu}_{\sigma}F^{c\sigma}_{\mu}f^{abc}}{v M^2}
\end{align*}
\begin{align}
 O_4 = \frac{H D^{\alpha}F^{a}_{ \alpha\nu}D_{\beta}F^{a \beta\nu}}{v M^2} \; O_5 = \frac{H F^{a}_{\alpha\nu}D^{\nu}D^{\beta}F^{a\alpha}_{\beta}}{v M^2}
\end{align}

Where $v$ is the vev of the Higgs particle and $M$ is the heavy scale at which the operators are 
generated. In the Standard Model, this would be the mass of the top quark.

Using the equations of motion, to this order in the large mass expansion, we can rewrite the 
last two operators in terms of quark bilinears as
\begin{align}
\frac{H D^{\alpha}F^{a}_{ \alpha\nu}D_{\beta}F^{a \beta\nu}}{v M^2} = & {}
\displaystyle\sum_{i=1,j=1}^{n_{f}}-g^2\frac{H \bar{q_i}\gamma_{\mu}T^a q_{i}\bar{q_j}\gamma^{\mu}T^a q_{j}}{v M^2} \\
\frac{H F^{a}_{ \alpha\nu}D^{\nu}D^{\beta}F^{a}_{\beta \alpha}}{v M^2} = & 
\displaystyle\sum_{i=1}^{n_{f}}-i g\frac{H F^{a}_{ \alpha\nu}D^{\nu}\bar{q_i}\gamma^{\alpha}T^a q_{i}}{v M^2}
\end{align}
where $n_f$ is the number of quark flavors. One can immediately see that the first operator will 
not contribute to any event involving three partons at this order: one needs to have a four 
partonic event for it to contribute, or higher loop corrections. It is further advantageous to 
decompose the operator $O_2$ as
\begin{align}
\frac{H D_{\alpha}F^{a}_{ \mu\nu}D_{\alpha}F^{a \mu\nu}}{v M^2} &=
-\frac{H \partial_{\alpha}\partial^{\alpha} (F^{a}_{\mu\nu}F^{a \mu\nu})}{2 v M^2}
+ 4\frac{H F^{a}_{\alpha\nu}D^{\nu}D^{\beta}F^{a\alpha}_{\beta}}{v M^2}
- 2\frac{HF^{a\mu}_{\nu}F^{b\nu}_{\sigma}F^{c\sigma}_{\mu}f^{abc}}{v M^2}\nonumber \\
&= -\frac{m_h^2}{2 M^2}O_1+4 O_5-2 O_3
\end{align}
where I have made use of the fact that on-shell the total momentum flowing through the $FF$ 
factor in the first operator must be negative that of the Higgs, if all particles are considered 
incoming. Thus the partial derivatives reduce to an over-all factor of $m_h^2$. 
The relation itself follows from expanding out one of the partial derivatives in the 
first operator and then making use of the Bianchi identity. With this decomposition, it is 
necessary to only calculate the one loop corrections to $O_5$ and $O_3$, as the $O_1$ amplitudes 
are well known \cite{Schmidt:1997wr}. 

\section{Methods}
Typically calculating one loop corrections to higher dimensional operators can be daunting due to the high rank
tensor integrals encountered. However, with the use of spinor helicity methods 
and generalized unitarity, one can tame the difficulty of such calculations in the
case of straight QCD and super Yang-Mills theories \cite{Bern:1994cg,Bern:1994zx}. Further, the
end results and intermediate steps exhibit a simplicity of form generally obscured by traditional methods of evaluation. 
The simplicity of analytic results come from using gauge invariant and onshell calculational techniques (and the
simple basis of master integrals at one loop). As such there is no reason to suppose that higher dimensional gauge 
invariant lagrangians ought to yield any more complicated scattering amplitudes. The only subtlies that can
arise are from the rational parts of the one-loop amplitudes.  If one takes strictly
four dimensional cuts, there are then undetermined rational contributions to the amplitude. Suitable methods for 
determining these rational contributions to amplitudes, such as on-shell recursion, 
typically break down for amplitudes at loop level. Specifically, there is a contribution 
from infinity when taking the large z limit of the amplitude for many choices of the onshell deformation of the amplitude. 
Often this large z contribution can be worked out from auxiliary shifts \cite{Berger:2006ci,Bern:2005cq,Bern:2005hs}.
For higher dimensional operators, this is even more problematic, as they have contributions from infinity even at tree-level \cite{ArkaniHamed:2008yf}.
In this regard the standard Higgs effective theory involving just the $HFF$ operator is the 
exception, due both its low dimensional order and its resemblance to pure QCD \cite{Berger:2006sh,Badger:2007si,Glover:2008ffa,Dixon:2009uk}. 

Thus for my purposes, it is advantageous to make use of d-dimensional unitarity \cite{Bern:1995db,Bern:1996ja,Anastasiou:2006jv,Anastasiou:2006gt}
to calculate the cuts, as by calculating cuts in d-dimensions (that is, the phase space of the cut is d-dimensional), 
one can determine the full analytic structure of the amplitude as long as the intermediate states are massless.
Also, cutting in d-dimensions, one can still make use of spinor integration to 
perform the tensor reduction \cite{Britto:2007tt,Britto:2006fc,Britto:2005ha}. This method is 
more direct than the standard Passarino-Veltmann reduction (particularly for many powers of 
loop momentum in the numerator), as the whole process of reduction to scalar integrals amounts 
to the extraction of residues in the spinor variables. These residues correspond directly
to the coefficients of scalar integrals. Furthermore, working directly with
spinor variables, the simple analytic structure becomes manifest.

\section{Conventions}
There are only two types of three partonic events in the Higgs production process: those
involving two quarks (or quark and anti-quark) and a gluon, or three gluon events. For each
operator we present the helicity partial amplitudes following the conventions of
so that a full amplitude has either of the two forms:
\begin{align}
M &= C*\frac{g}{2 v}tr(T^a[T^b,T^c])m(1^{\lambda_{1}},2^{\lambda_{2}},3^{\lambda_{3}})\\
M &= C*\frac{g}{2 v}T^a_{i_q i_{\overline{q}}} m(1^{\lambda_{1}}_{\overline{q}},2^{\lambda_{2}}_q,3^{\lambda_{3}})
\end{align}
where $C$ would be the appropriate matching coefficient, and the index $i$ denotes the incoming
momenta with helicity $\lambda_i$. Subscripts $q$ or $\overline{q}$ denote quark or
anti-quark momenta, and will be suppressed where appropriate.

The amplitudes are presented here unrenormalized, so that the correct counterterms must still be
added. I give the minimal helicity amplitudes, other helicity configurations either vanish, or
are obtainable from parity or charge conjugation. We use Weyl spinors  $|p\pm\rangle $ and denote their 
products as $\langle p_i-|p_j+\rangle =\langle ij\rangle $ and $\langle p_i+|p_j-\rangle =[ij]$ with the
QCD sign conventions $\langle ij\rangle [ji]= \langle i|p_j|i] =2p_i \cdot p_j= S_{ij}$. 
For spinor chains, one writes $\langle i|a|j] = \langle i a \rangle[a j]=a_{\mu}\langle i|\gamma^{\mu} |j]$ and
$\langle i|a b|j\rangle = \langle i a \rangle [a b]\langle b j \rangle$ with $a$ and $b$ null vectors. 
The $SU(N_c)$ fundamental matrices are normalized as $tr(T^aT^b)=\delta_{a b}$ and $[T^a,T^b]=i\sqrt{2} f^{abc}T^c$.

I regularize keeping the external states 4-dimensional, and extending
the loop momenta in $4-2\epsilon$ dimensions, and the tensor algebra in $4-2\epsilon\delta$
dimensions. Then one can obtain results in  t'Hooft-Veltmann scheme \cite{'tHooft1972189} by 
taking $\delta \rightarrow 1$ or Four Dimensional Helicity scheme 
\cite{PhysRevLett.66.1669,Bern:1991aq} with $\delta \rightarrow 0$.

\section{Constructing the Cut Integrands}
To construct the integrands for the cut amplitudes, one must be careful to make sure that
they are valid in d-dimensions for the cut momenta. Taking the naive 4-dimensional
tree amplitudes written in terms of helicity spinors and extrapolating to d-dimensions
can lead to errors in the rational terms \cite{Bern:1994cg,Mahlon:1993fe,Mahlon:1993si}.
One can then either build tree amplitudes with the external states to be used in the loop 
integration to be d-dimensional, recover the rational terms from on-shell recursion or 
collinear constraints \cite{Bern:1994zx} (taking all cuts as strictly four dimensional), 
or build specialized Feynman rules for the rational part \cite{Ossola:2008xq}. This last course of action 
applied to a non-renormalizable theory requires knowledge of the rational parts generated 
by higher rank tensor integrals, which, while straightforward to obtain, have not been classified.
Then a matching calculation is needed to extract out the loop level rational contributions to the Feynman rules.

Thus in a theory with a non-renormalizable power counting, one can either use Feynman diagrams
or an extension of off-shell recursion \cite{Berends:1987me}, or try to tackle the 
on-shell recursion with a boundary term. Given the low number of partons, it is easiest 
to build the integrands from Feynman diagrams using color-ordered Feynman rules. Each side of the 
cut amplitude is taken to be the sum of Feynman diagrams in Feynman gauge contributing to the tree 
level process on that side of the cut. Any gluon lines from the tree amplitude running across the 
cut are then truncated, and the polarization sum is replaced by the metric tensor. This leaves only the 
ghost diagrams to be calculated separately. To check gauge invariance of the final result, I calculate the cut 
for several choices of the reference momentum of the external polarizations. The final
result then is independent of these choices.

\section{Spinor Integration}
One obtains for a general two particle cut:

\begin{align}
A_{1-loop}|_{K-cut} = \sum_{\lambda_{1} \lambda_{2}}&\int d^DL_1\delta^{+}(L_1^2)d^DL_2\delta^{+}(L_2^2) \delta^{D}(L_2-L_1-K)\nonumber\\ &A^{tree}_L(L_1^{\lambda_{1}},p_i,...,p_j,-L_2^{\lambda_2})A^{tree}_R(L_2^{-\lambda_{2}},p_k,...,p_l,-L_1^{-\lambda_1})
\end{align}

where $K=p_i+...+p_j=-(p_k...+p_l)$, $\lambda_i$ are the polarizations of the intermediate states,
and the momenta is considered to be incoming on all amplitudes. Further, one can integrate over 
the $L_1$ momentum in favor of $L_2$ and write

\begin{align}
A_{1-loop}|_{K-cut} = \sum_{\lambda_{1} \lambda_{2}} 
\int& d^DL\delta^{+}(L^2)\delta^{+}((L-K)^2)\nonumber\\ &A^{tree}_L((L-K)^{\lambda_{1}},....,-L^{\lambda_{2}})A^{tree}_R(L^{-\lambda_{2}},...,(K-L)^{-\lambda_1})
\end{align}

To make use of spinor integration to perform the scalar reduction, one further decomposes
the integration variable into 4-dimensional ($\overline{l}$) and $-2\epsilon$ dimensional, 
($\mu$) pieces, $L=\overline{l}+\mu$. (Since the $-2\epsilon$ dimensions are considered
space-like, one interprets $\mu \cdot \mu$ as $-\mu^2$, with $\mu^2$ an explicitly positive quantity. 
So $(L+P)^2=(\overline{l}+P)^2-\mu^2$, where $P$ is any four dimensional vector.) As long as 
the external states are four dimensional (so $\mu$ can only form a product with itself)
the integration over the angles in the $-2\epsilon$ dimensions may be performed, 
so that the integration measure becomes:
\begin{align}
\int d^DL\delta^{+}(L^2)\delta^{+}((L-K)^2) &= \frac{(4\pi)^{\epsilon}}{\Gamma(-\epsilon)}\int d\mu^2(\mu^2)^{-1-\epsilon}d^4\overline{l}\delta^+(\overline{l}^2-\mu^2)\delta^+((\overline{l}-K)^2-\mu^2)
\end{align}

As explained in \cite{Anastasiou:2006gt}, given a fixed time-like vector $K$ (taken as the cut momentum),
one can further decompose the 4-dimensional part of the integration as an integration over a 
single parameter $z$ and a null vector $l$ as $\overline{l}=l+zK$.
\begin{align}
\int d^4\overline{l} &= \int d^4l\delta(l^2)dz(2 l \cdot K)
\end{align}
This decomposition is what allows for spinor integration to be used in d-dimensions.
The integration over the null vector can be written in terms of spinor variables \cite{Cachazo:2004kj},
$\int d^4l\delta(l^2) = \int_0^{\infty} t dt \int \langle l dl\rangle [l dl]$. Scalar products $2l \cdot P$ then
satisfy $2l \cdot P= t\langle l|P|l]$. The two delta functions from the cuts allow one to integrate over 
the $t$ and $z$ integrals. The $\mu$ integral never need formally be evaluated, as scalar integrals in the 
basis have unique analytic forms in $\mu$.

One is left then only considering the integration in the spinor products $\int \langle l dl\rangle [l dl]$.
Specifically, a term in the expression for the cut 1-loop amplitude will transform as,
\begin{align}
\int& d^DL\delta^{+}(L^2)\delta^{+}((L-K)^2)\frac{\prod_{i=1}^{m}2L \cdot P_{i}}{\prod_{j=1}^{k}(L-K_{j})^2} =\nonumber\\ &\frac{(4\pi)^{\epsilon}}{\Gamma(-\epsilon)}\int d\mu^2(\mu^2)^{-1-\epsilon} \int \langle ldl\rangle [ldl] \frac{(K^2)^{m-k+1}}{\langle l|K|l]^{m-k+2}}\frac{\prod_{i=1}^{m}\langle l|R_i|l]}{\prod_{j=1}^{k}\langle l|Q_j|l]}
\end{align}
The $R_i$ and $Q_j$ are $\mu^2 (=_{df}\frac{K^2}{4}u)$ dependent four momenta derived from $P_i$ and $K_j$,
\begin{align}
R_i &= (\sqrt{1-u})P_i + \frac{(1-\sqrt{1-u})P_i \cdot K}{K^2}K\\
Q_j &= -(\sqrt{1-u})K_j + \frac{K_j^2-(1-\sqrt{1-u})K_j \cdot K}{K^2}K
\end{align}

The actual reduction proceeds by using a form of the Schouten Identity to partial fraction the spinor products:
\begin{align}
\frac{[al]}{[bl][lc]}=\frac{[ab]}{[bc][bl]}+\frac{[ac]}{[bc][lc]}
\end{align}
Thus the spinor products $[l \bullet]$ may be eliminated in favor of $\langle l \bullet\rangle $ up to a
holomorphic anomaly (or vice versa) \cite{Britto:2005ha,Britto:2006sj}. 

%

The final spinor integration reduces to evaluating the holomorphic anomaly and the residues 
of the integrand in the spinor variable $|l\rangle $. These residues can neatly be organized into 
the coefficients of the scalar integrals in the decomposition of an amplitude. The required 
formulas for the coefficients can be found in \cite{Britto:2007tt} and were derived in \cite{Britto:2006fc}.

The scalar integral basis is built from pentagon, box, triangle, and bubble scalar integrals defined in higher
dimensions. More precisely, one finds the coefficient for an integral of the form:

\begin{align}
I^{D+2N}_{n}=\int d^{D+2N}L \frac{1}{L^2(L-k_1)^2...(L-k_{n-1})^2}
\end{align}

where $n\leq 5$, $N$ is some integer, $k_i$ a sum over a subset of the external momentum and $D=4-2\epsilon$. 
One generates these higher dimensional integrals from the spinor integration, 
as the residues from the integration are in general polynomials in $\mu^2$ multiplying the
analytic expression for the cut integral when $N=0$. These powers of $\mu^2$ can then be 
reinterpreted as part of the measure of a scalar integral in another spacetime dimension \cite{Bern:1995db,Britto:2008sw},
giving rise to the higher dimensional integrals. 

\section{Calculating with $HFFF$}
Here I present the calculation of the all positive helicity gluon one-loop correction to the $HFFF$
operator. The contribution from this operator is particularly easy to evaluate as
the four dimensional cuts completely determine the amplitude, and due symmetry of the helicity
configuration, a cut in only one channel needs to be evaluated in order to find the full contribution.

The cut in momentum $K=p_1+p_2$ gives the integral for the process as simply:
\begin{align}
I=-\int d^dL \frac{6[12]^2(\langle 1|L|3][13]+\langle 2|L|3][23])}{2(L-p_2)^2M^2}\delta^+(L^2)\delta^+((L-K)^2)
\end{align}
At this point, the integral is simple enough that one could just evaluate by
using the Cutkovsky rules to convert it back into a loop integral, and either Passarino-Veltmann
reduce, or reduce via Feynman parameters. One then evaluates the cut of the resulting
expressions. 

However, for illustration purposes, I go ahead and use the full machinery of the spinor integration.
Following the above steps for setting up the spinor integration, one finds
\begin{align}
I=-\frac{6[12]^2}{M^2}\frac{(4\pi)^{\epsilon}}{\Gamma(-\epsilon)}\int d\mu^2(\mu^2)^{-1-\epsilon} \int \langle ldl\rangle [ldl] \frac{(K^2)}{\langle l|K|l]^{2}}\frac{\langle l|R_{13}|l][13]+\langle l|R_{23}|l][23]}{\langle l|Q|l]}
\end{align}
where $Q=-(\sqrt{1-u})p_2-(1-\sqrt{1-u})\frac{K}{2}$, $R_{ij}=(\sqrt{1-u})P_{ij}+(1-\sqrt{1-u})P_{ij}\cdot K \frac{K}{K^2}$,
and $P_{ij}^{\nu} =\frac{1}{2}\langle i|\gamma^{\nu} |j]$. At this point one can directly apply the
formulas given in \cite{Britto:2007tt}.

\subsection{Triangle Coefficient}
There is only one triangle integral contributing to the amplitude in this cut.
The coefficient itself is given by the residue of the poles of $\langle l|KQ|l \rangle$. These poles
are easily found to be given by the Weyl spinors of the momenta $P_+ = (\sqrt{1-u})p_1$ and $P_- = -(\sqrt{1-u})p_2$.
That is $|P_+ \rangle= |1\rangle$,$|P_+]= (\sqrt{1-u})|1]$, and $|P_- \rangle= |2\rangle$,$|P_-]=-(\sqrt{1-u})|2]$.
The residue is given by 
\begin{align}
C_{Tri} =& -\lim_{\tau \rightarrow 0}\frac{6[12]^2}{2\sqrt{1-u}M^2}\frac{1}{\langle P_+ P_- \rangle}\frac{d}{d\tau}(\langle P_+-\tau P_-|R_{13}Q|P_+-\tau P_-\rangle [13]\nonumber\\ 
&+\langle P_+-\tau P_-|R_{23}Q|P_+-\tau P_-\rangle [23]+(P_+ \leftrightarrow P_-))\\
=& -\langle 12 \rangle[12] \frac{6[13][23][12]}{M^2}
\end{align}

\subsection{Bubble Coefficient}
The bubble coefficient is also straightforward to find. Here there is a auxiliary null vector $\eta$ introduced when
evaluating the holomorphic anomaly. Often there can be subtlies when $\eta$ is fixed such that it coincides with
another pole in the integrand. However, here one can take it to be $p_3$ with no issues and proceed normally.
\begin{align}
C_{Bubble} =& -\frac{6[12]^2}{M^2}(\frac{2\eta \cdot K}{\langle \eta|K|\eta]}\frac{\langle l|R_{13}K|l\rangle [13]+\langle l|R_{23}K|l\rangle [23]}{\langle l|QK|l\rangle} |_{|l\rangle \rightarrow K|\eta]}\nonumber\\
&- \frac{ 1}{\sqrt{1-u}}(\frac{\langle P_+|\eta|P_+]\langle P_+|([13]R_{13}+[23]R_{23}) K|P_+\rangle}{\langle P_+|K|P_+] \langle P_+|\eta K|P_+\rangle}\nonumber\\
&- \frac{\langle P_-|\eta|P_-]\langle P_-|([13]R_{13}+[23]R_{23}) K|P_-\rangle}{\langle P_-|K|P_-] \langle P_-|\eta K|P_-\rangle}))\\
=& \frac{12[13][23][12]}{M^2}
\end{align}

\subsection{Full Amplitude}
Combining the bubble and the triangle together, and using the symmetric helicity configuration, one has,
\begin{align}
m^1_{HFFF}(1^+,2^+,3^+)= \frac{6[13][23][12]}{M^2}(&2 I^{4-2\epsilon}_2(p_1+p_2)+2p_1 \cdot p_2 I^{4-2\epsilon}_3(p_1,p_2,(p_3+p_h))\nonumber\\
&+ 2 I^{4-2\epsilon}_2(p_1+p_3)+2p_1 \cdot p_3 I^{4-2\epsilon}_3(p_1,p_3,(p_2+p_h))\nonumber\\
&+ 2 I^{4-2\epsilon}_2(p_2+p_3)+2p_2 \cdot p_3 I^{4-2\epsilon}_3(p_2,p_3,(p_1+p_h)))
\end{align}
where $I^D_n$ is the $D$ dimensional $n$ point scalar integral.

\section{Results}

\begin{figure}
\includegraphics[width=30 mm, height=20 mm]{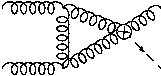}
\caption{A contributing diagram with an insertion of the $HFFF$ operator.}
\end{figure}

\begin{figure}
\includegraphics[width=30 mm, height=20 mm]{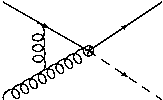}
\caption{A contributing diagram with an insertion of the $HFDJ$ operator.}
\end{figure}

For reference I first give the tree level $HFF$ helicity partial amplitudes.
\begin{align}
m^0_{HFF}(1^+,2^+,3^+) &= \frac{m_h^4}{\langle 12\rangle \langle 23 \rangle \langle 31 \rangle}\\
m^0_{HFF}(1^+,2^+,3^-) &= -\frac{[12]^4}{[12][23][31]}\\
m^0_{HFF}(1^-_{\overline{q}},2^+_q,3^+) &= \frac{[23]^2}{[12]}
\end{align}

At tree level we have for the $HFFF$ operator:
\begin{align}
m^0_{HFFF}(1^+,2^+,3^+) &= -6 \frac{[12][23][31]}{M^2}\\
m^0_{HFFF}(1^+,2^+,3^-) &= 0
\end{align}

At one loop I obtain:
\begin{align}
m^1_{HFFF}(1^+,2^+,3^+) &= m^0_{HFFF}(1^+,2^+,3^+)r_{\Gamma}N_c \frac{\alpha_s}{4\pi}(
(\frac{1-4\epsilon}{\epsilon^2(1-2\epsilon)})
((\frac{-m_h^2}{-S_{12}})^\epsilon +(\frac{-m_h^2}{-S_{23}})^\epsilon
+(\frac{-m_h^2}{-S_{13}})^\epsilon))+O(\epsilon)\\
m^1_{HFFF}(1^+,2^+,3^-) &= 0+O(\epsilon)
\end{align}

The $HFDJ$ operator has no tree level coupling to gluons, and at loop level couples through a 
quark loop, giving
\begin{align}
m^1_{HFDJ}(1^+,2^+,3^+) &= -m^0_{HFFF}(1^+,2^+,3^+)n_{f}\frac{\alpha_s}{4\pi}+O(\epsilon)\\
m^1_{HFDJ}(1^+,2^+,3^-) &= -m^0_{HFF}(1^+,2^+,3^-)n_{f}\frac{\alpha_s}{4\pi}(\frac{S_{13}S_{23}}{3 M^2 S_{12}})+O(\epsilon)
\end{align}
where $n_{f}$ is the number of light quarks. Note that $m^0_{HFF}$ is the tree level partial amplitude for the $HFF$ operator.

Finally, I give the amplitudes involving two quarks and a gluon. The $HFFF$ has no tree level coupling
to quarks at this order, and at loop level,
\begin{align}
m^1_{HFFF}(1^-_{\overline{q}},2^+_q,3^+) = m^0_{HFF}(1^-_{\overline{q}},2^+_q,3^+)N_c \frac{\alpha_s}{4\pi} \delta(\frac{S_{12}}{6 M^2})+O(\epsilon)
\end{align}

The amplitudes for the $FHDJ$ operator are

\begin{align}
m^0_{HFDJ}(1^-_{\overline{q}},2^+_q,3^+) &= -\langle 12\rangle [23]^2\\
m^1_{HFDJ}(1^-_{\overline{q}},2^+_{q},3^+) &= m^0_{HFDJ}(1^-_{\overline{q}},2^+_{q},3^+)(\frac{\alpha_s}{4\pi} r_\Gamma)(N_c U_1+\frac{1}{N_c}U_2)\\
U_1 &= \frac{1}{\epsilon^2}((\frac{-m_h^2}{-S_{13}})^\epsilon+(\frac{-m_h^2}{-S_{23}})^\epsilon) -\frac{1}{\epsilon(1-2\epsilon)}(\frac{11}{12}(\frac{-m_h^2}{-S_{13}})^\epsilon +(\frac{-m_h^2}{-S_{23}})^\epsilon)\nonumber\\
 &- \frac{S_{12}}{3 S_{23}}-\frac{S_{13}}{6 S_{23}}-\frac{3\delta+16}{18}+O(\epsilon)\\ 
U_2 &= \frac{1}{\epsilon^2}(\frac{-m_h^2}{-S_{12}})^\epsilon -\frac{3}{2\epsilon(1-2\epsilon)}(\frac{-m_h^2}{-S_{12}})^\epsilon +\frac{1+\delta}{2}+O(\epsilon)
\end{align}

Note that 
\begin{align}
r_\Gamma = (\frac{4\pi\mu^2_{renorm}}{m_h^2})^\epsilon\frac{\Gamma(1+\epsilon)\Gamma(1-\epsilon)^2}{\Gamma(1-2\epsilon)}.
\end{align}

It is straightforward to check that the presented amplitudes preproduce the correct anomalous dimensions where known (the $HFFF$ operator) 
\cite{Gracey:2002rf} and the correct scheme dependence as given in \cite{Catani:1996pk}.

\section{Conclusions}
I have presented the virtual partonic amplitudes necessary to calculate the subleading terms in the heavy mass expansion 
at two-loops to the Higgs $p_t$ spectrum, by using unitarity based calculational methods on operators with a
non-renormazible power counting. While this is a straightforward generalization of its established use in 
renormalizable theories, the power counting complicates a naive application of the
methods. Though such power suppressed operators are often negligible for phenomonology, in some sectors 
they are the simplest way of estimating higher loop effects at a higher mass scale. Having machinery that allows
for simple calculations with such operators is therefore desireable, and unitarity and spinor helicity methods provides
such methods. In a forth coming publication I will use these amplitudes to determine the 
subleading effects on the $p_t$ differential cross-section.

\section{Acknowledgements}
The author would like to thank Ira Rothstein for useful discussions, Amber Jain and Zvi Bern for reading the manuscript. Diagrams were made with Jaxodraw \cite{Binosi:2003yf}. Work supported by DOE contracts 
DOE-ER-40682-143 and DEAC02-6CH03000 with partial support from LHC Fellows NSF grant PHY-0705682.

\vspace{-.4cm}

\bibliographystyle{h-physrev}
\bibliography{higgs_three_partons_paper}

\newpage

\end{document}